\documentclass[3p,times,twocolumn]{elsarticle}

\usepackage{ecrc}

\volume{00}

\firstpage{1}

\journalname{Nuclear Physics B Proceedings Supplement}

\runauth{}

\jid{nuphbp}

\jnltitlelogo{Nuclear Physics B Proceedings Supplement}

\usepackage{amssymb}
\usepackage[figuresright]{rotating}

\begin{document}

\begin{frontmatter}



\dochead{}

\title{Recent results on charm physics at BESIII}


\author{Hai-Bo Li (for BESIII Collaboration)}
%\thanks{on behalf of BESIII Collaboration}
\address{Institute of High Energy Physics, Yuquan Road 19-2, Beijing 100049, China}

\begin{abstract}
About 2.9 fb$^{-1}$ data set had been collected with the BESIII
detector at the BEPCII collider.  Taking advantages of data near
open-charm threshold, we present preliminary results on leptonic and
semileptonic charm-meson decays, as well as result on the FCNC
process $D^0 \rightarrow \gamma \gamma$  from BESIII experiment.
High precision charm data will enable us to validate Lattice QCD
calculations at the few percent level. These can then be used to
make precise measurements of CKM elements, $V_{cd}$ and $V_{cs}$,
which are useful to improve the accuracy of test of the CKM unitary.
%The prospects on rare and forbidden charm decays at BESIII are also
%reviewed.
\end{abstract}


\end{frontmatter}

% \linenumbers

\section{Introduction}
\label{sec:intro}

The charm physics potential at open-charm threshold plays important
role on quark flavor physics~\cite{linpb-2006}. The open-charm
program at BESIII includes studies of leptonic, semileptonic,
hadronic charm-meson decays, searches for neutral D mixing, CP
violations, and rare and forbidden charm decays, which are sensitive
to physics beyond standard Model (SM). The quark mixing parameters
are fundamental constants of the SM. They determine the nine
weak-current quark coupling elements of the CKM matrix~\cite{ckm}.
Studies of the semileptonic or pure-leptonic decays of D mesons are
preferred way to determine the CKM elements, $|V_{cs}|$ and
$|V_{cd}|$, since the strong interaction binding effects are
parameterized by form factors or decay constants that are
calculable, for example, by lattice QCD (LQCD) and QCD sum rules. On
the other hand, $|V_{cs}|$ and $|V_{cd}|$ are tightly constrained
when CKM matrix is assumed to be unitary. Therefore, measurements of
charm semileptonic or pure-leptonic decay rates rigorously test
theoretical prediction of the D meson semileptonic form factors or
decay constants. High precision predictions of QCD will then remove
road blocks for many weak and flavor physics measurements, such as
in B decays for determinations of $|V_{ub}|$ and neutral B mixing
parameters.

Many of the measurements related to charm decays  are also
accessible to the B-factories and Super-B factories. What are the
advantages to running at the open charm threshold at the BESIII
experiment?  The BESIII experiment will not be able to compete
Super-B factories in statistics on charm physics, especially on the
rare and forbidden decays of charm mesons. However, data taken at
charm threshold still have powerful advantages over the data at
$\Upsilon(4S)$, which we list here~\cite{linpb-2006,gibbons}: 1)
Charm events produced at threshold are extremely clean; 2) the
measurements of absolute branching fraction can be made by using
   double tag events, which was first applied by the MARKIII
   Collaboration at SPEAR~\cite{markiii}. The produced $\psi(3770)$
   in our sample decays into a pair of $D\bar{D}$. Reconstructing
   one of the $D$ mesons with know exclusive hadronic modes while
   looking for decays of the other
    $D$ mesons would allow us to reconstruct final states with
    neutrinos;
3) signal/Background is optimum at threshold; 5) Quantum coherence
allow simple~\cite{gronau} and complex~\cite{asner} methods to
measure $D^0\bar{D}^0$ mixing parameters,  direct $CP$ violation and
strong phase differences.

In this talk, I report  preliminary results on charm meson decays
based on a sample collected at the BEPCII with the BESIII
detector~\cite{besiii}. This sample was collected at
$\sqrt{s}=3.773$ GeV with an integrated luminosity of 2.9 fb$^{-1}$
in which the background levels at the open charm threshold is
expected to be substantially lower than that at the $\Upsilon(4S)$
peak at $B$ factories.

\section{Purely leptonic $D$ decay }
\label{sec:lepton}

 With a sample of 2.9 fb$^{-1}$ taken at open-charm threshold, BESIII experiment
 measures the decay branching fraction for $D^+ \rightarrow \mu^+
 \nu_{\mu}$ and extracts decay constant $f_{D^+}$~\cite{gang-charm2012}.
 Decay constant characterizes
 the strong-interaction physics at the quark-annihilation vertex. In a fully leptonic decay,
 they parameterize all of our essential theoretical limitations. Decay constant for $B$ case also
appears in the evaluation of box diagrams, and limit theoretical
precision in calculating the neutral $B$ meson mixing. Thus, lack of
knowledge of the $B^0$ and $B_s$ decays constants limits the
usefulness of precise measurements of $B^0-\bar{B}^0$ and $B_s -
\bar{B}_s$ oscillations. These mixing data are our best sources of
information on the CKM matrix elements $V_{td}$ and $V_{ts}$, which
are difficult to measure directly in top decay. The leptonic decay
of charm meson presents an opportunity to check LQCD results for
decay constants against precision measurements.
\begin{figure}[htbp]
  \begin{center}
    \includegraphics[width=0.4\textwidth]{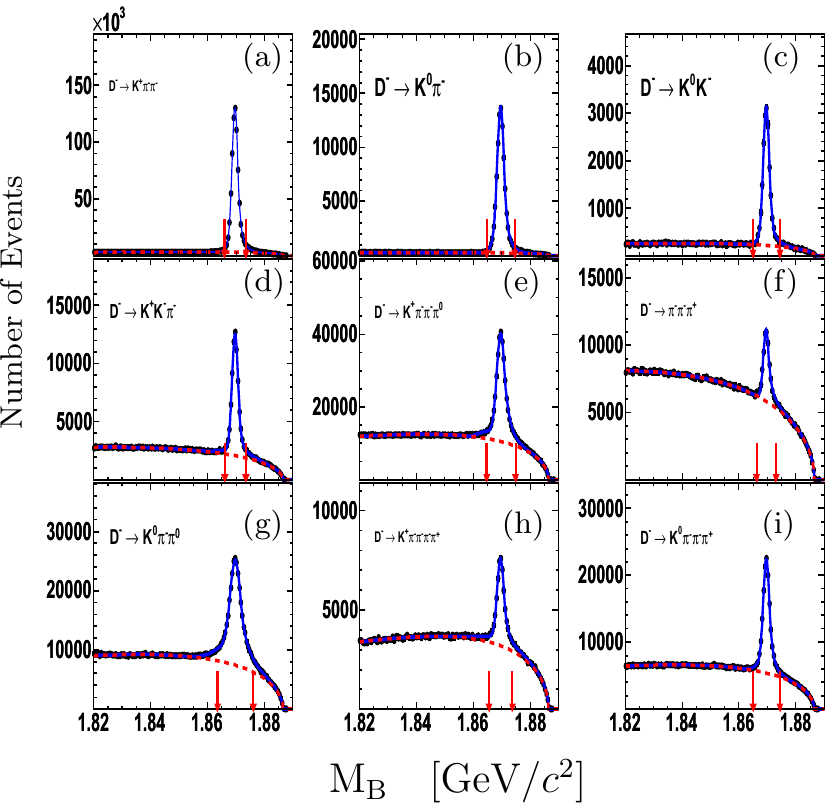}
    \caption{Distributions of the beam energy constraint masses for $D^-$ hadronic tags used for the
$D^+ \rightarrow \mu^+ \nu$ analysis. Modes in panels (a)-(i) are
$D^-\rightarrow K^+\pi^-\pi^-$, $K_s\pi^-$, $K_s K^-$,
$K^+K^-\pi^-$, $K^+\pi^-\pi^-\pi^0$, $\pi^+\pi^-\pi^-$, $K_s
\pi^-\pi^0$, $K^+ \pi^-\pi^-\pi^-\pi^+$, and $K_s \pi^-\pi^-\pi^+$.}
    \label{leptonic-tag}
  \end{center}
\end{figure}

A ¡°tag¡± is simply a fully-reconstructed $D$ hadronic decays. A
sample of tagged events has greatly reduced background and
constrained kinematics, both of which aid studies of how the other
$D$ in the event decays. One can infer neutrinos from energy and
momentum conservation, allowing ¡°full¡± reconstruction of
(semi)leptonic $D$ decays. The typical tag rates per $D$ (not per
pair) are roughly 15\% and 10\% for $D^0$ and $D^+$, respectively.
For pure leptonic decay in this analysis, the singly tagged $D^-$
mesons are reconstructed in nine non-leptonic decay modes of
$D^-\rightarrow K^+\pi^-\pi^-$, $K_s\pi^-$, $K_s K^-$,
$K^+K^-\pi^-$, $K^+\pi^-\pi^-\pi^0$, $\pi^+\pi^-\pi^-$, $K_s
\pi^-\pi^0$, $K^+ \pi^-\pi^-\pi^-\pi^+$, and $K_s \pi^-\pi^-\pi^+$.
Mass peaks for the nine hadronic tag modes are shown in
Fig.~\ref{leptonic-tag}. A maximum likelihood fit to the mass
spectrum yields the number of the singly tagged $D^-$ events for
each of the nine modes. The total number of tagged $D^-$ events are
$1565953 \pm 2327$ $D^-$.

\begin{figure}[htbp]
  %\begin{center}
    \includegraphics[width=0.4\textwidth]{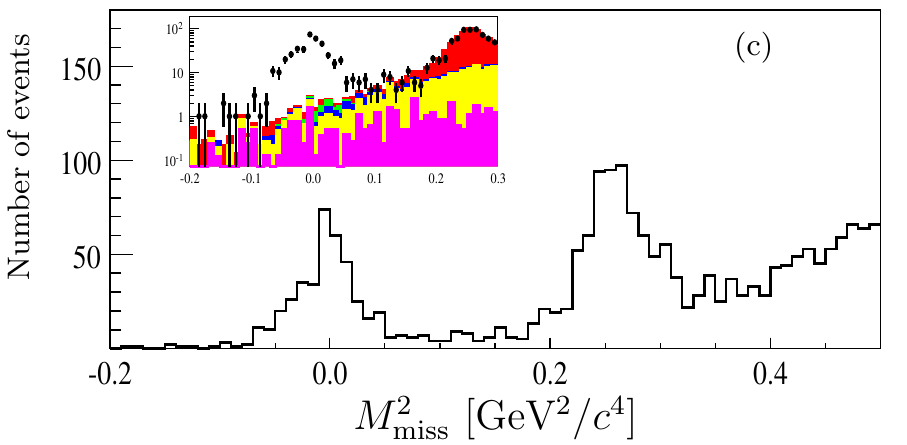}
    \caption{BESIII missing-mass-squared plot for $D^+ \rightarrow
\mu^+\nu$. The insert shows the signal region on a vertical log
scale, where dots with error bars are for the data, histograms are
for the simulated backgrounds from $D^+ \rightarrow K_L \pi^+$
(red), $D^+\rightarrow \pi^+ \pi^0$ (green), $D^+ \rightarrow
\tau^+\nu$ (blue) and other decays of $D$ mesons (yellow) as well as
from $e^+e^- \rightarrow$ non-$D\bar{D}$ decays (pink).}
    \label{lepton-sig}
 % \end{center}
\end{figure}

The chosen signal variable for the $\mu^+\nu$ decay is the
calculated square of the missing-mass of any undetected decay
products, shown in Fig.~\ref{lepton-sig}; this should of course peak
at $M^2_{\nu}= 0$ for signal events. The power of $D$-tagging is
evident in the clean, isolated signal peak. In
Table~\ref{background-lepton}, sources of background modes are
summarized including $K_L \pi^+$, $\pi^+\pi^0$, $\tau\nu$ and other
components.
\begin{table}[htbp]
\begin{center} \caption{Sources of background events for $D^+ \rightarrow
\mu\nu$.} \label{background-lepton}
\begin{tabular}{@{}ll}
\hline\hline
Source mode    & Number of events  \\
\hline
$D^+\rightarrow K_L\pi^+$ & $7.9\pm0.8$\\
$D^+\rightarrow \pi^+\pi^0$   &  $3.8\pm0.5$ \\
$D^+\rightarrow \tau^+\nu$   & $6.9\pm0.7$ \\
Other $D$ decays & $17.9\pm0.1$ \\
ISR to $\psi^\prime$ and $J/\psi$ & $0.2\pm0.2$ \\
$\psi(3770) \rightarrow Non-D\bar{D}$ & $0.9\pm0.4$\\
$e^+e^- \rightarrow q\bar{q}$ & $8.2\pm1.4$ \\
$e^+e- \rightarrow \tau^+\tau^-$ & $1.9\pm0.5$ \\
\hline Total  & $47.7\pm 2.3$  \\ \hline\hline
\end{tabular}
\end{center}
\end{table}
After subtracting the expected number of background events, about
$377.3 \pm 20.6$ signal events for $D^+ \rightarrow \mu^+ \nu$ decay
are retained, where the error is statistical. The overall efficiency
for observing the decay $D^+ \rightarrow \mu^+\nu$ is determined  to
be $63.82\%$ by analyzing full Monte Carlo (MC) simulation events of
$D^+ \rightarrow \mu^+\nu$ {\it versus} $D^-$ tags. Therefore, we
obtain the preliminary branching fraction to be
\begin{equation}
\mathcal{B}(D^+\rightarrow \mu^+\nu )= (3.74\pm 0.21\pm0.06)\times
10^{-4} , \label{eq:lepton}
\end{equation}
where the first error is statistical and the second systematic. This
measured branching fraction is consistent within error with world
average of $\mathcal{B}(D^+ \rightarrow \mu^+ \nu) =
(3.82\pm0.33)\times 10^{-4}$~\cite{pdg2012}, but with more
precision. The decay constant $f_{D^+}$ is then obtained using $1040
\pm 7$ fs as the $D^+$ lifetime and 0.2256 as
$V_{cd}$~\cite{pdg2012}. Our preliminary result is
\begin{equation}
f_{D^+}= (203.91 \pm 5.72 \pm 1.97)\, MeV, \label{eq:decay}
\end{equation}
where the first errors are statistical and the second systematic
arising mainly from the uncertainties in the measured branching
fraction (1.7\%), the CKM matrix element $V_{cd}$ (0.3\%), and the
lifetime of the $D^+$ meson (0.7\%)~\cite{pdg2012}. The total
systematic error is 1.0\%. In Table~\ref{comparison-lepton}, the
BESIII preliminary results are compared with
CLEO-c~\cite{cleo-c-lepton} and LQCD predictions~\cite{hpqcd} for
the decay constant. The preliminary result from BESIII is the best
number in the context of the SM, and remains consistent with LQCD.
\begin{table}[htbp]
\footnotesize
\begin{center}
\caption{Comparison of results for $D^+ \rightarrow \mu\nu$ and
decay constant from BESIII~\cite{gang-charm2012},
CLEO-c~\cite{cleo-c-lepton} and LQCD prediction~\cite{hpqcd}.}
\label{comparison-lepton}
\begin{tabular}{@{}lll}
\hline\hline
Model  & $\mathcal{B}(D^+\rightarrow \mu^+\nu)\times 10^{-4}$  & $f_{D^+}$ (MeV)  \\
\hline
BESIII & $3.74\pm 0.21\pm0.06$ & $203.91 \pm 5.72 \pm 1.97$\\
CLEO-c~\cite{cleo-c-lepton} &  $3.82\pm 0.32\pm0.09$ & $205.8 \pm
8.5 \pm 2.5$  \\\hline \hline Average &$3.76\pm0.18$ & $204.5\pm5.0$
\\ \hline HPQCD~\cite{hpqcd} & - & $213\pm4$ \\\hline \hline
\end{tabular}
\end{center}
\end{table}

\section{Semileptonic $D$ decays: $D^0 \rightarrow K^- e^+ \nu$ and $\pi^- e^+\nu$ }
\label{sec:lepton}

One of the best ways to measure magnitudes of CKM elements is to use
semileptonic decays since they are far simpler to understand than
hadronic decays and the decay width is $\sim$ $|V_{cq}|^2$. On the
other hand, measurements using other techniques have obtained useful
values for $V_{cs}$ and $V_{cd}$~\cite{ckmcharm}, and thus
semileptonic $D$ decay measurements are a good laboratory for
testing theories of QCD. For a $D$ meson decaying into a single
hadron ($h$), the decay rate can be written exactly in terms of the
four-momentum transfer defined as:
\begin{equation}
q^2= (p^{\mu}_D -p^{\mu}_h)^2 = m^2_D +m^2_h -2E_h m_D.
\label{eq:decay}
\end{equation}
For decays to pseudoscalar mesons and ¡°virtually massless¡±
leptons, the decay width is given by:
\begin{equation}
\frac{d\Gamma(D\rightarrow P e^+\nu)}{dq^2}=\frac{G^2_F|V_{cq}|^2
p^3_P}{24\pi^3}|f_+(q^2)|^2, \label{eq:decay-semi}
\end{equation}
where $p_P$ is the three-momentum of pseudoscalar meson in the $D$
rest frame, and $f_+(q^2)$ is a ¡°form-factor,¡± whose normalization
must be calculated theoretically, although its shape can be
measured.

\begin{figure}[htbp]
  \begin{center}
    \includegraphics[width=0.2\textwidth]{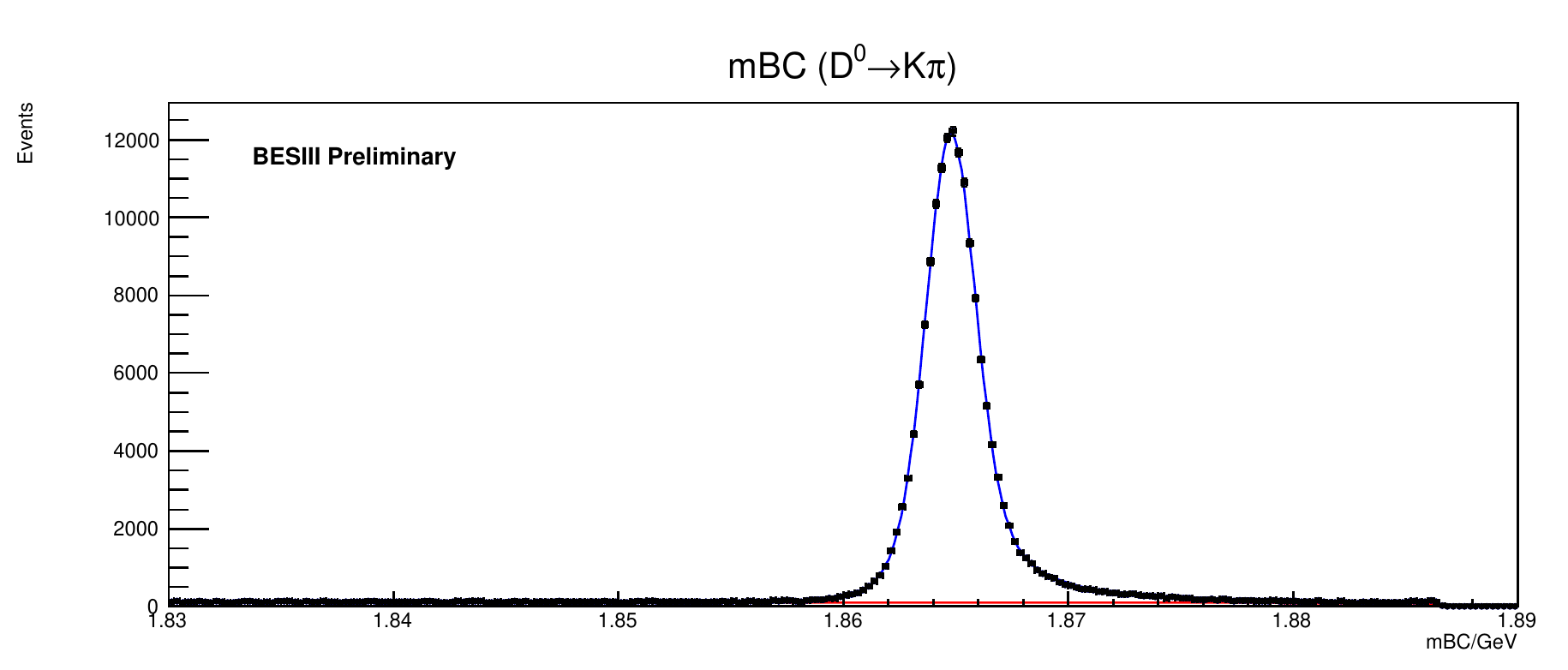}
    \includegraphics[width=0.2\textwidth]{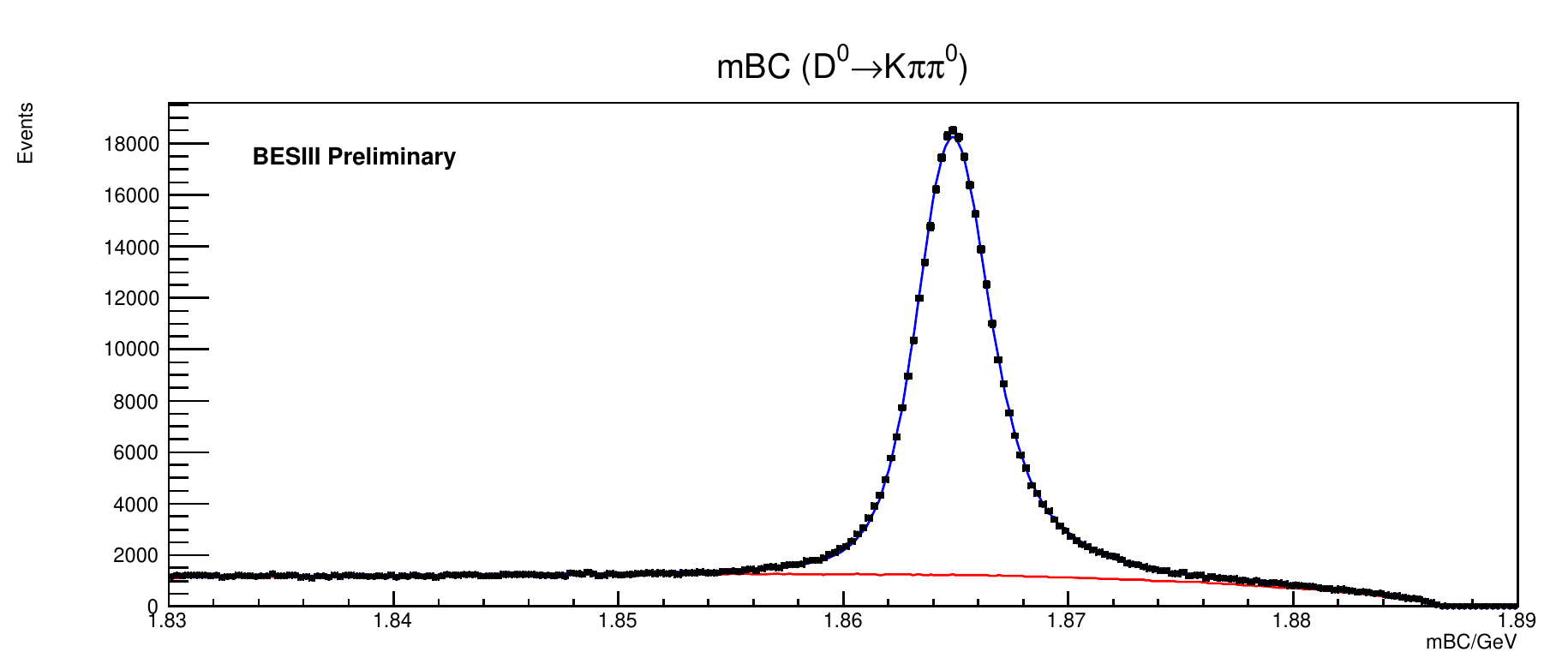}
    \includegraphics[width=0.2\textwidth]{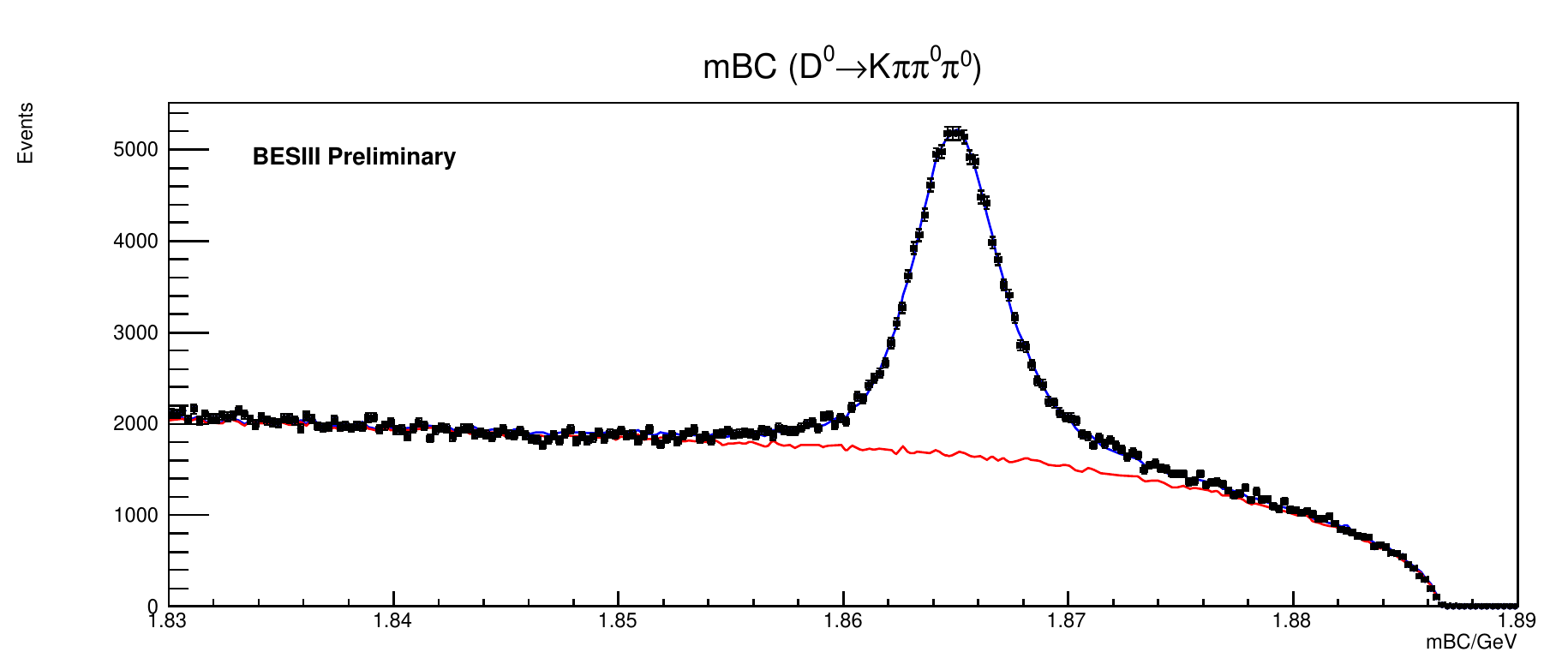}
    \includegraphics[width=0.2\textwidth]{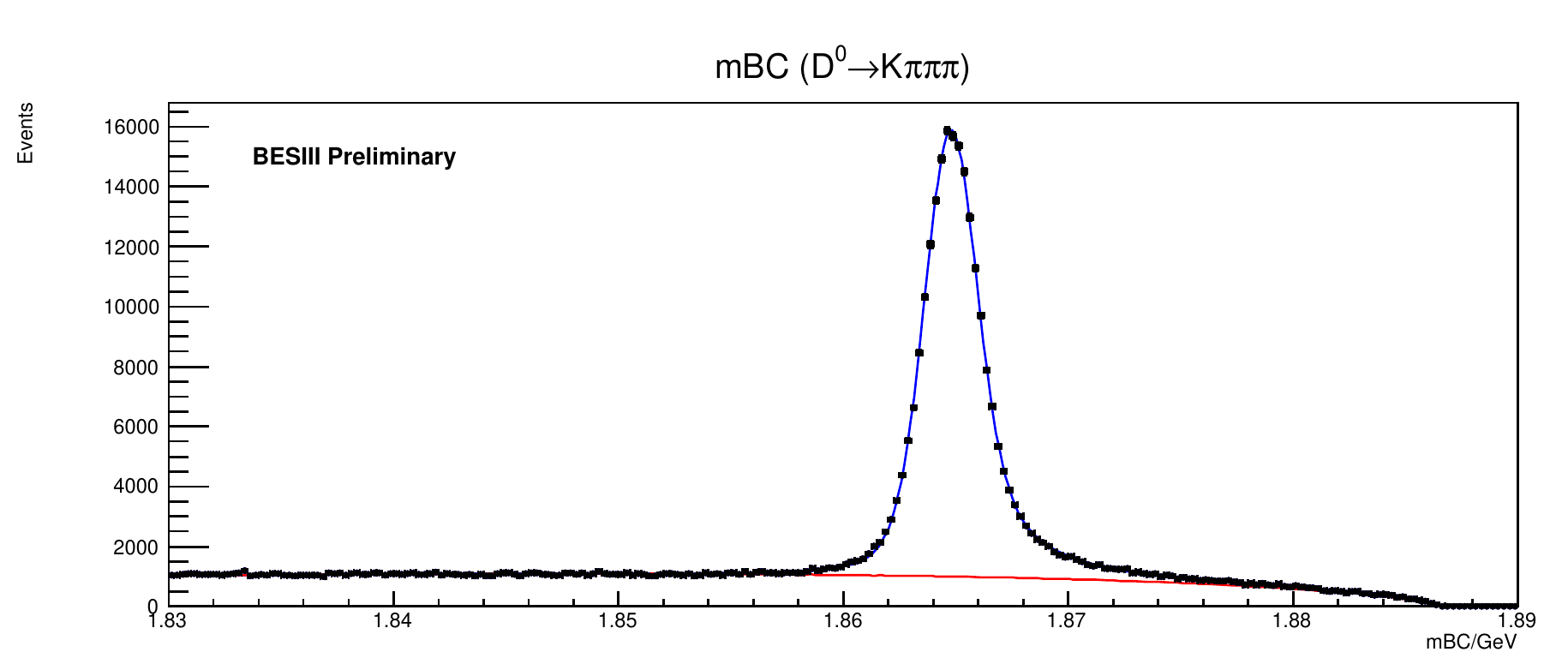}
    \caption{Distributions of the beam energy constraint masses for $D^0$ hadronic tags used for the
$D^0 \rightarrow K e \nu$ and $\pi e\nu$  analyses. Modes in panels
 are $D^0\rightarrow K^-\pi^+$, $K^-\pi^+\pi^0$,
$K^-\pi^+\pi^0\pi^0$ and $K^-\pi^+\pi^-\pi^+$.}
    \label{semi-tag}
  \end{center}
\end{figure}

\begin{figure*}[htbp]
  \begin{center}
    \includegraphics[width=0.3\textwidth]{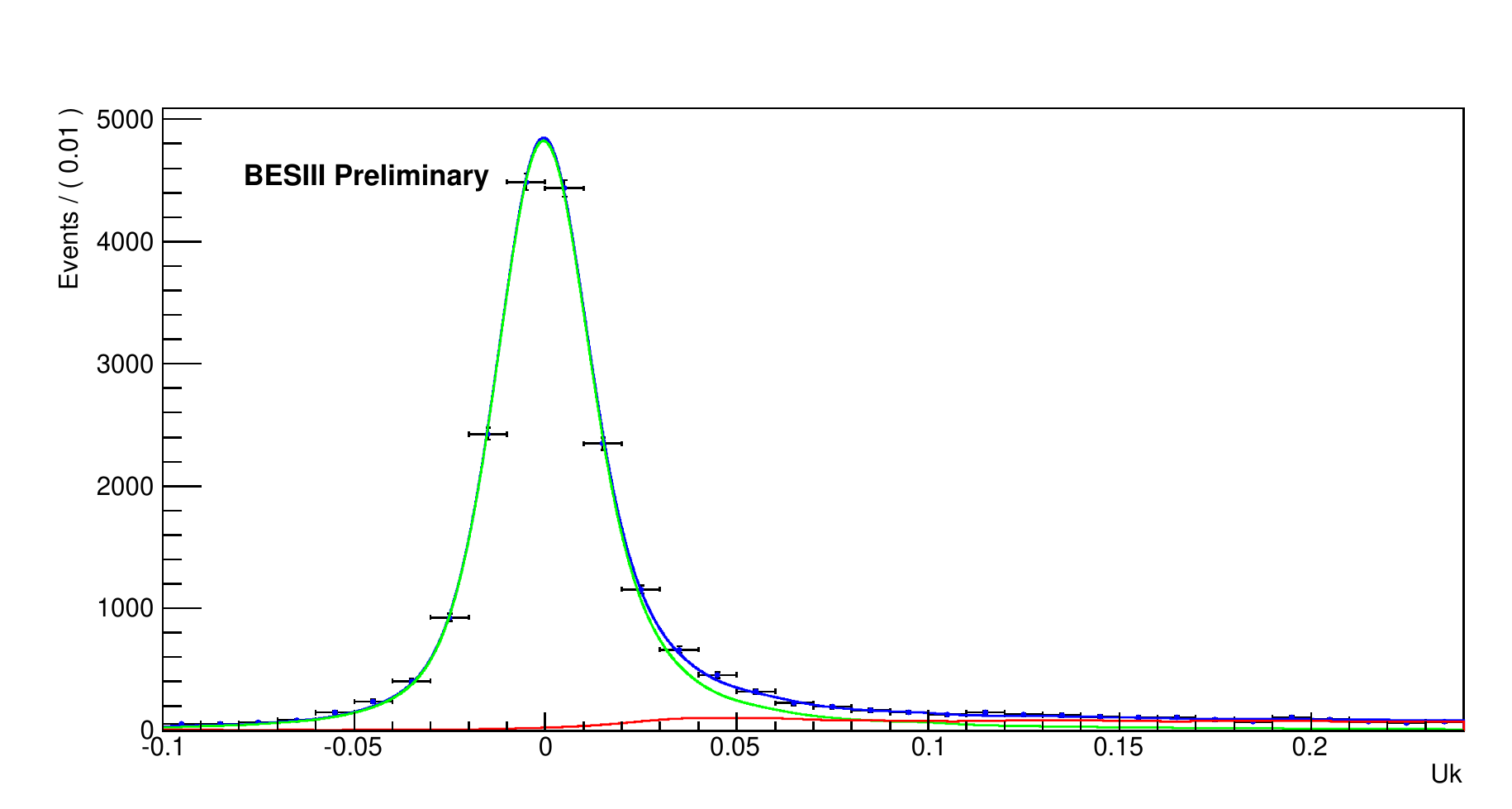}
    \includegraphics[width=0.3\textwidth]{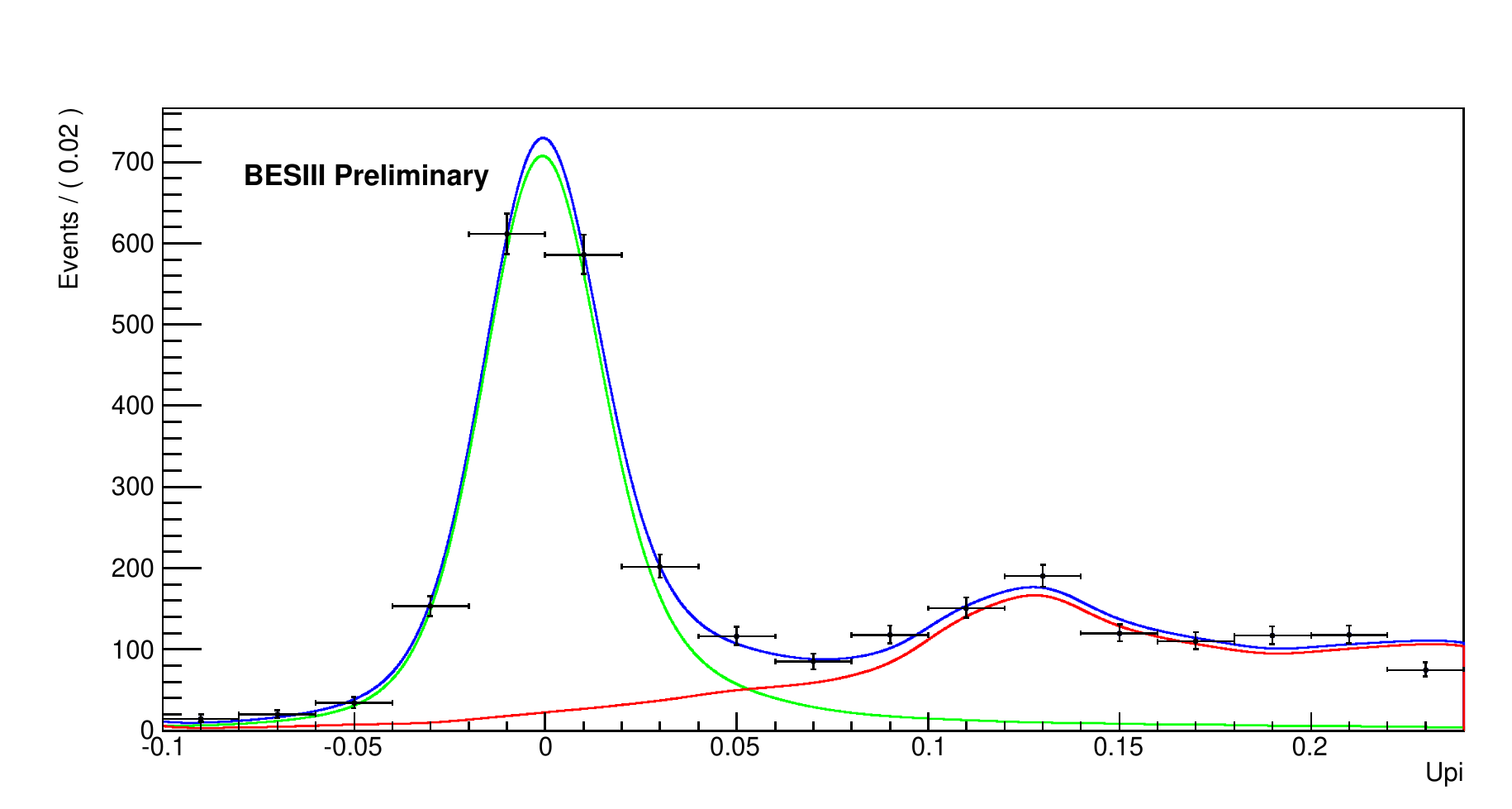}
    \caption{$U$ distributions of $\bar{D}^0 \rightarrow K^+e^-\nu$(left) and $\bar{D}^0 \rightarrow
     \pi^+ e^- \nu$(right). Blue, green,
and red curves are the total fit, signal fit, and background fit,
respectively.}
    \label{fig:umiss}
  \end{center}
\end{figure*}

The BESIII experiment has taken about 2.9 fb$^{-1}$ data at
open-charm threshold during the 2010 and 2011 data runs. Using
one-third of the data, a partially-blind analysis has been done with
the $D^0 \rightarrow  Ke\nu$ and $D^0\rightarrow \pi e\nu$ decays.
Using the double tag technique, several hadronic $D$ decays are
fully reconstructed at first. The following four hadronic $D$ decays
are used: $D^0 \rightarrow K^-\pi^+$, $K^-\pi^+\pi^0$,
$K^-\pi^+\pi^0\pi^0$ and $K^-\pi^+\pi^-\pi^+$. Mass peaks for the
four hadronic tag modes are shown in Fig.~\ref{semi-tag}.
After hadronic $D^0$ tags are found, we reconstruct signal decay for
the other $\bar{D}^0$. The signal events with a missing $\nu$ are
inferred using the variable $U=E_{miss} - |P_{miss}|$, similar to
missing mass square, where "miss" here refers to the missing energy
or momentum. Figure.~\ref{fig:umiss} shows the $U$ distributions and
fit projections for the decays of $\bar{D}^0 \rightarrow K^+ e^-
\nu$ and $\bar{D}^0 \rightarrow \pi^+ e^- \nu$.

\begin{figure*}[htbp]
  \begin{center}
    \includegraphics[width=0.3\textwidth]{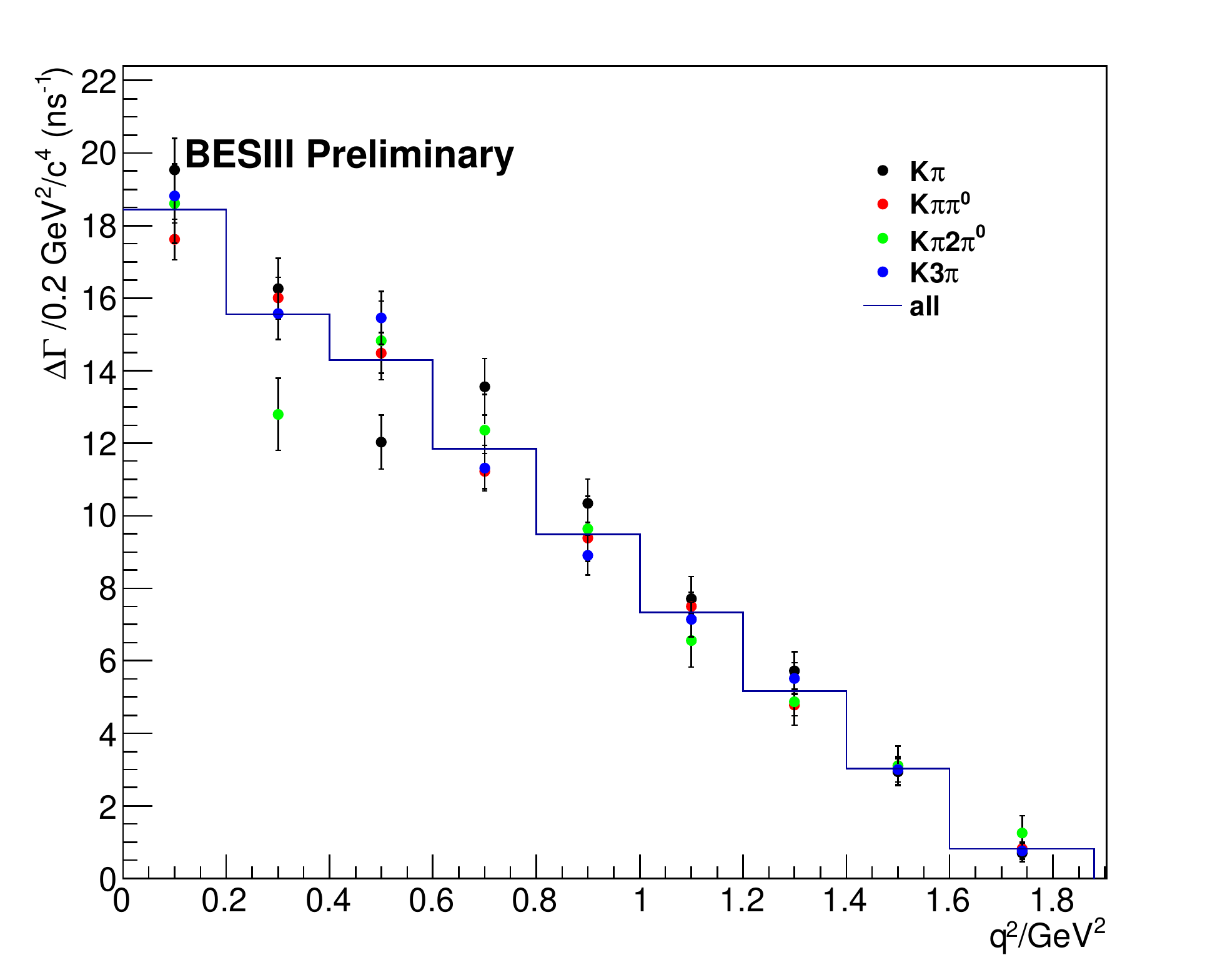}
    \includegraphics[width=0.3\textwidth]{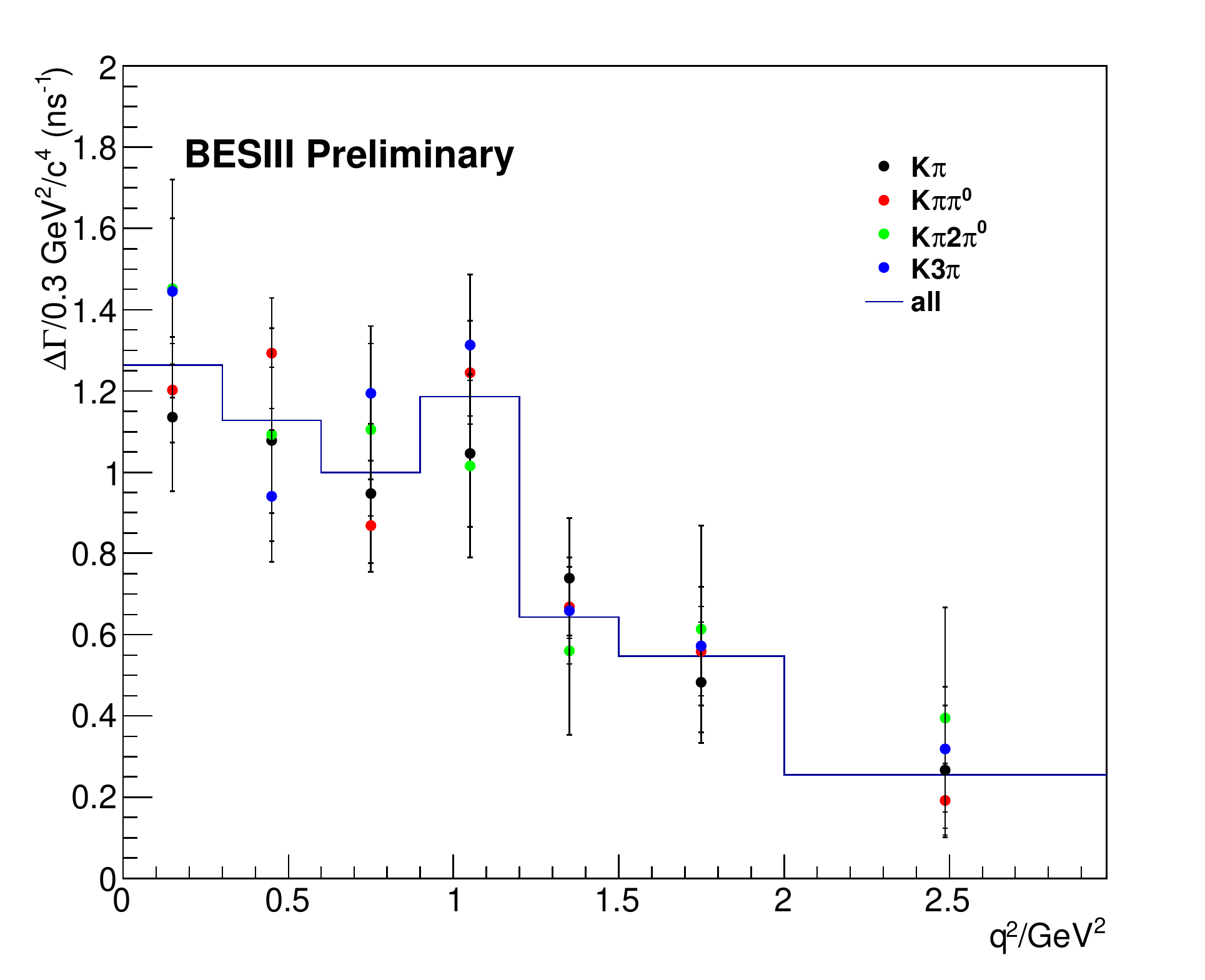}
    \caption{Partial decay rates measurement using individual tag modes (points) and all
tag modes combined (histogram) for decay of $\bar{D}^0\rightarrow
K^+ e^- \nu$ (left) and $\bar{D}^0 \rightarrow \pi^+ e^-\nu$
(right).}
    \label{fig:partialbr}
  \end{center}
\end{figure*}

\begin{figure}[htbp]
  \begin{center}
    \includegraphics[width=0.5\textwidth]{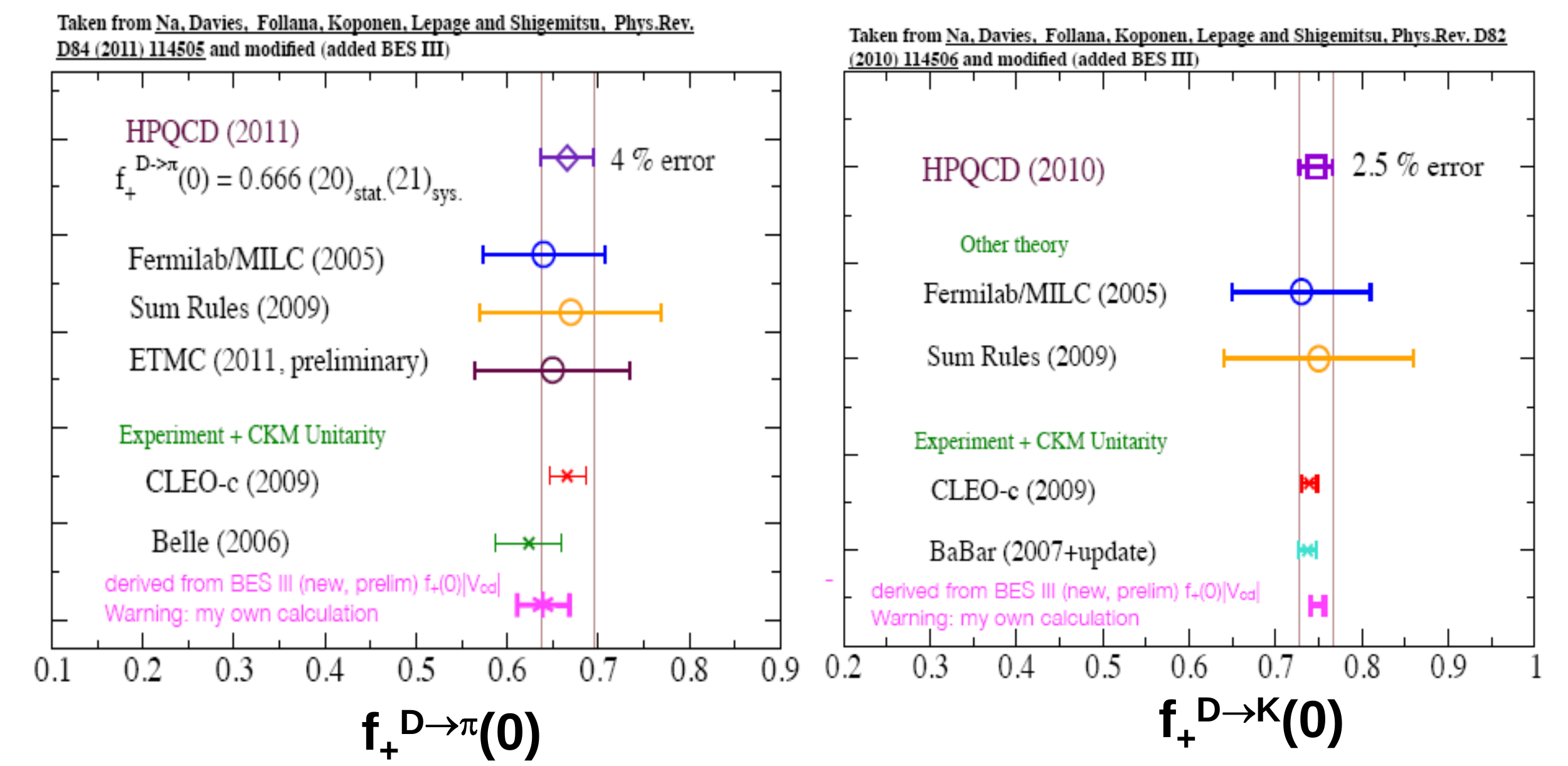}
    \caption{Comparison of $f_+(0)$ from experiments and theory. Left plot is for $f^{D\rightarrow
    \pi}_+(0)$,
and right plot is for $f^{D\rightarrow K}_+(0)$. Note that BESIII
result from $D^0$ decays only, while CLEO-c use both $D^0$ and $D^+$
decays~\cite{rad-fpcp2012}. }
    \label{fig:f0}
  \end{center}
\end{figure}

Given the signal yields obtained from fitting $U$ distributions and
signal efficiencies obtained from signal Monte Carlo, the absolute
branching fractions are obtained. Preliminary results of branching
fractions are listed in Table~\ref{semi-br}, and comparisons with
results from PDG2012~\cite{pdg2012} and CLEO-c
results~\cite{cleo-c-semi} are also made. In order to measure form
factor, partial decay rates are measured in different $q^2$ bins.
$\bar{D}^0 \rightarrow K^+ e^-\nu$ candidates are divided into nine
$q^2$ bins, while $\bar{D}^0 \rightarrow \pi^+ e^- \nu$ candidates
are divided into seven $q^2$ bins. Signal yields in each $q^2$ bin
are obtained by fitting $U$ distributions in that $q^2$ range. Using
an efficiency matrix {\it versus} $q^2$, obtained from Monte-Carlo
simulation, and combining with tag yields and tag efficiencies, the
partial decay rates are obtained, as shown in
Fig.~\ref{fig:partialbr}.
\begin{table}[htbp]
\footnotesize
\begin{center}
\caption{Branching fraction measurement using 923 pb$^{-1}$ of
$\psi(3770)$ data from BESIII experiment, and comparisons with
results from CLEO-c~\cite{cleo-c-semi} and PDG2012~\cite{pdg2012}.}
\label{semi-br}
\begin{tabular}{@{}lll}
\hline\hline
Experiment  & $\mathcal{B}(\bar{D}^0\rightarrow K^+e^-\nu)(\%)$  & $\mathcal{B}(\bar{D}^0\rightarrow \pi^+ e^-\nu)(\%)$  \\
\hline
BESIII & $3.542\pm 0.030\pm0.067$ & $0.288 \pm 0.008 \pm 0.005$\\
CLEO-c~\cite{cleo-c-semi} &  $3.50\pm 0.03\pm0.04$ & $0.288 \pm
0.008 \pm 0.003$
\\ PDG2012~\cite{pdg2012} &$3.55\pm0.04$ & $0.289\pm0.008$ \\ \hline
 \hline
\end{tabular}
\end{center}
\end{table}
\begin{figure}[htbp]
  \begin{center}
    \includegraphics[width=0.3\textwidth]{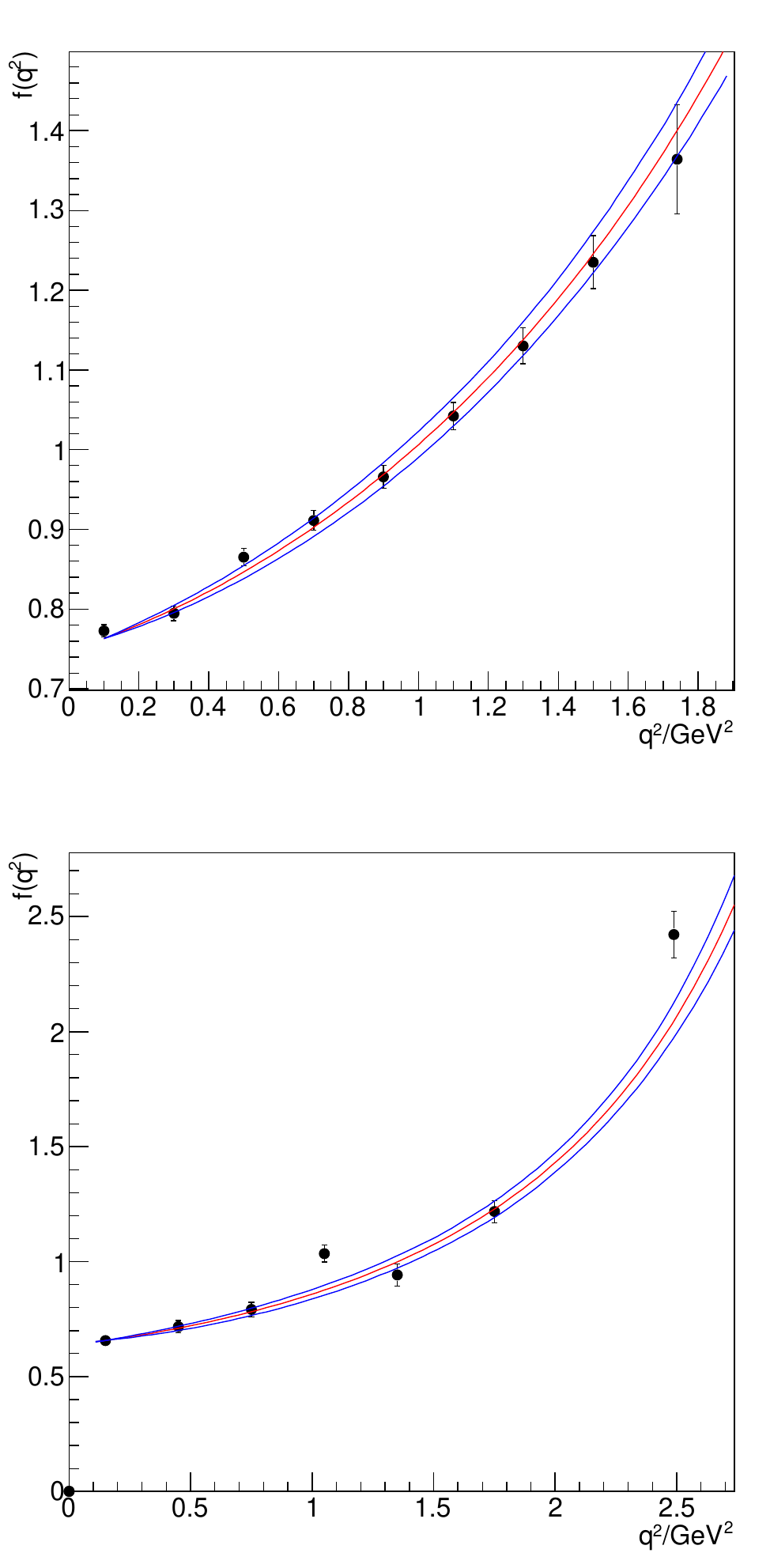}
    \caption{Plot of $f_+(q^2)$ from data with Lattice QCD prediction. Black points are from data with
statistical error only, red curve is theoretical prediction, and
blue curves are theoretical variations within statistical error
only. The top plot is for  $\bar{D}^0\rightarrow  K^+ e^- \nu$  and
the bottom plot is for $\bar{D}^0 \rightarrow \pi^+ e^-\nu$
(bottom).}
    \label{fig:form-values}
  \end{center}
\end{figure}
The values of $q^2$-dependent form factors in each $q^2$ bin can be
extracted from the measured partial decay rates as shown in
Fig.~\ref{fig:form-values}. These data can be fitted with different
parameterizations of the form factors, and the fit can distinguish
between form factors parameterizations. In general, one may express
the form factors in terms of a dispersion relation, an approach that
has been well established in the literature (see, for example,
Ref.~\cite{form-factors} and references therein):
\begin{eqnarray}
f_+(q^2)&=& \frac{f_+(0)}{1-\alpha}\frac{1}{(1-q^2/m^2_{pole})}+\nonumber\\
 &&\frac{1}{\pi}\int^{\infty}_{(m_D +m_P)^2}
\frac{Im(f_+(t))}{t-q^2-i\epsilon} dt,
 \label{eq:form-pole}
\end{eqnarray}
where $m_{pole}$ is the mass of the lowest lying $(q_i \bar{q}_f)$
meson with the appropriate quantum numbers: for $D \rightarrow
Ke\nu$ it is $D_s^{*+}$ and for $D\rightarrow \pi e \nu$ it is
$D^{*+}$, the parameter $\alpha$ gives the relative contribution
from the vector pole at $q^2 =0$, $m_D$ is the mass of the $D$
meson, and $m_P$ is the mass of the final state pseudoscalar meson.
The integral term can be expressed in terms of an infinite
series~\cite{form-factors}. Typically it takes only a few terms to
describe the data. Three different parameterizations of the form
factor $f_+(q^2)$ are considered. The first parameterization, known
as the simple pole model, is dominated by a single
pole~\cite{form-factors1}; the second parameterization is known as
the modified pole model~\cite{form-factors1}; the third
parameterization is known as the series
expansion~\cite{form-factors}. Thus minimized $\chi^2$ fits are
employed to extract the values of $f_+(0)|V_{cd(s)}|$ using each of
the parameterizations. The preliminary results for
$f_+(0)|V_{cd(s)}|$  are shown in Table~\ref{tab:form-factor}. With
$|V_{cd}| =0.2252$ ($|V_{cs}| = 0.97345$)~\cite{pdg2012} and  BESIII
new results (3 par. series)~\cite{liu-charm2012}, we extract the
values for $f^{D\rightarrow \pi}_+(0)$ and $f^{D\rightarrow
K}_+(0)$, and results are compared with other experiments and
theoretical calculations as shown in
Fig.~\ref{fig:f0}~\cite{rad-fpcp2012}.
\begin{table}[htbp]
\begin{center}
\caption{Results of $f_+(0)|V_{cd(s)}|$ from individual form factor
fits; statistical and systematic uncertainties on the least
significant digits are shown in parentheses. Results from
CLEO-c~\cite{cleo-c-semi} are compared.} \label{tab:form-factor}
\newcommand{\cc}[1]{\multicolumn{2}{c}{#1}}
\begin{tabular}{@{}ccc}
\hline\hline
  &  \cc{$f_+(0)|V_{cd(s)}|$}  \\
\hline & BESIII  & CLEO-c \\ \hline
3 par. Series &  & \\
$\bar{D}^0\rightarrow K^+e^-\nu$ &  0.729(8)(7) & 0.726(8)(4)\\
$\bar{D}^0\rightarrow \pi^+e^-\nu$ & 0.144(5)(2)&  0.152(5)(1)\\
\hline
2 par. Series &  & \\
$\bar{D}^0\rightarrow K^+e^-\nu$ &  0.726(6)(7) & 0.717(6)(4)\\
$\bar{D}^0\rightarrow \pi^+e^-\nu$ &0.140(4)(2) & 0.145(4)(1) \\
\hline
Modified pole &  & \\
$\bar{D}^0\rightarrow K^+e^-\nu$ &  0.725(6)(7) & 0.716(6)(4)\\
$\bar{D}^0\rightarrow \pi^+e^-\nu$ & 0.140(3)(2)&  0.145(4)(1)\\
\hline
Simple pole &  & \\
$\bar{D}^0\rightarrow K^+e^-\nu$ &  0.729(5)(7) & 0.720(5)(4)\\
$\bar{D}^0\rightarrow \pi^+e^-\nu$ & 0.142(3)(1)&  0.146(3)(1)\\
\hline
 \hline
\end{tabular}
\end{center}
\end{table}

\section{Rare charm decays at BESIII: preliminary results on $D^0 \rightarrow \gamma \gamma$ }
\label{sec:rare }

Searches for rare-decay processes have played an important role in
the development of the SM. Short-distance flavor-changing neutral
current (FCNC) processes in charm decays are much more highly
suppressed by the GIM mechanism than the corresponding down-type
quark decays because of the large top quark mass. Observation of
FCNC decays $D \rightarrow h l^+l^-$ and $D \rightarrow  l^+l^-$
could therefore provide indication of new physics or of unexpectedly
large rates for long-distance SM processes like $D \rightarrow h V$,
$V \rightarrow l^+l^-$, with real or virtual vector meson $V$.

With 2.9 fb$^{-1}$ data at $\psi(3770)$ peak, BESIII search for $D^0
\rightarrow \gamma \gamma$ decay which must be produced by FCNC.
From the short distance contributions, the decay rate for $D^0
\rightarrow \gamma \gamma$ is predicted to be $3\times 10^{-11}$
~\cite{gg1,gg2,gg3}. However, the long distance contributions
significantly enhance the decay rate which is estimated to be $(1
-3)\times 10^{-8}$~\cite{gg2,gg3}. This decay could be enhanced by
new physics (NP) effects which lead to contributions at loop
level~\cite{gg4,gg5}. For instance, in the framework of the Minimal
Supersymmetric Standard Model (MSSM), the calculation shows that the
decay rate for $c \rightarrow u \gamma$
 transition could be $6 \times 10^{-6}$, which is one to two orders of
 magnitudes enhanced relative to the SM rate, by considering gluino exchange~\cite{gg4}.

Experimental searches for $D^0 \rightarrow \gamma\gamma$ were
performed by the CLEO~\cite{gg6} and BABAR~\cite{gg7} experiments
based on data samples collected at the $\Upsilon(4S)$ peak. They
found no significant signals. The latter experiment yields the most
stringent experimental upper limit to date on the $\mathcal{B}(D^0
\rightarrow \gamma \gamma)$, $2.2 \times 10^{-6}$ at 90\% confidence
level (C.L.).

\begin{figure}[htbp]
  \begin{center}
    \includegraphics[width=0.3\textwidth]{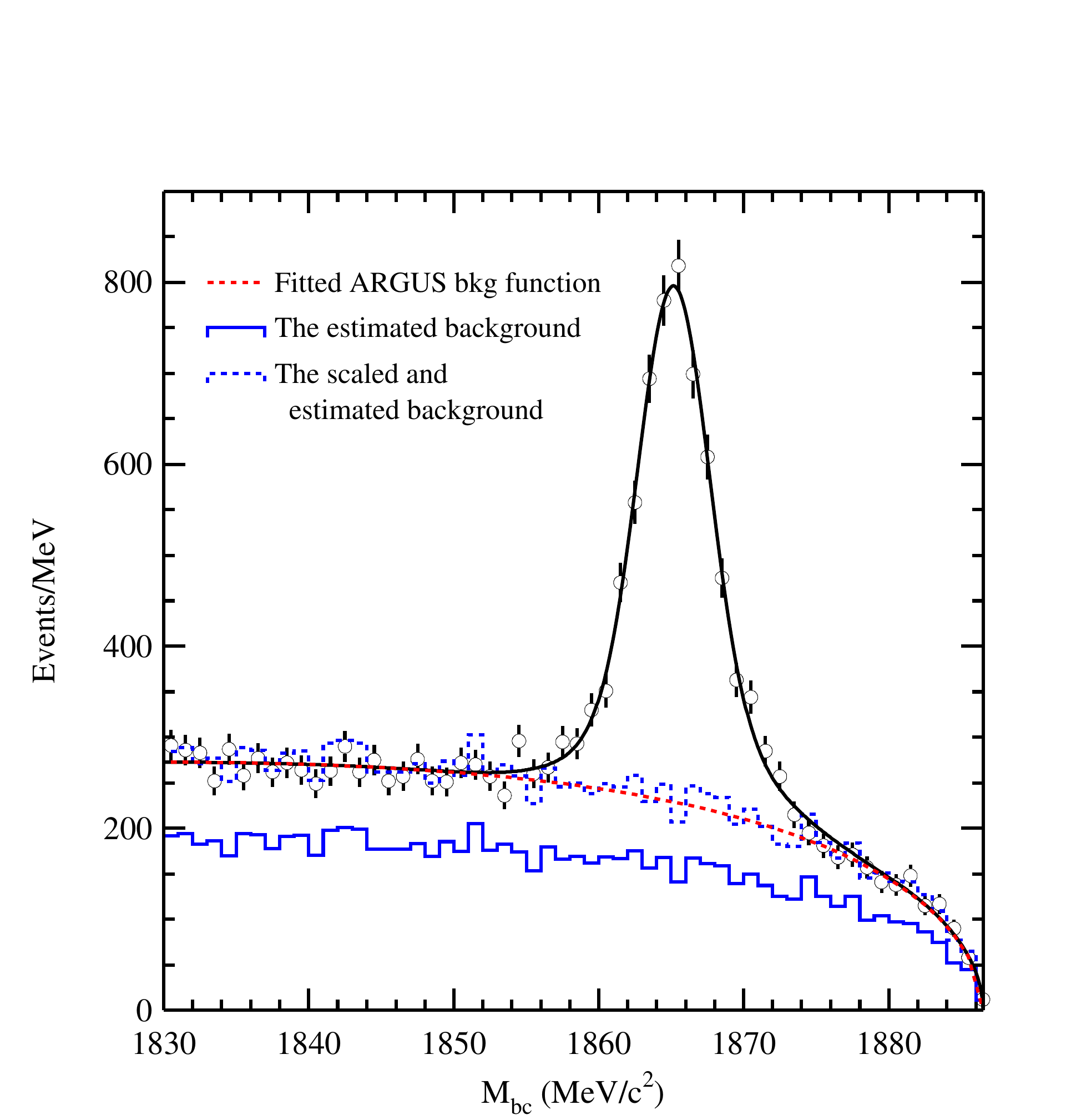}
    \caption{A fit to the $D^0 \rightarrow \pi^0\pi^0$ candidate $M_{bc}$
    distribution based on the $\psi(3770)$
data sample. The black points are data, the black-smooth curve
represents the overall fit (signal plus background), and the
red-dashed curve corresponds to the fitted background shape. The
blue-solid histogram represents the expected background shape and
size based on our MC samples while the blue-dotted histogram is a
fit to the data based on this expected MC-based background shape. }
    \label{fig:2pi0}
  \end{center}
\end{figure}

Since one of the major backgrounds in the analysis of $D^0
\rightarrow \gamma \gamma$ comes from $D^0 \rightarrow \pi^0\pi^0$,
we firstly measure the decay rate of $D^0\rightarrow \pi^0\pi^0$
using the same data sample. Then we use the measured $D^0\rightarrow
\pi^0\pi^0$ decay rate to do the normalization and background
estimation for $D^0 \rightarrow \gamma \gamma$ measurement.
Figure~\ref{fig:2pi0} shows the beam-constraint mass distribution of
the observed $\pi^0\pi^0$ signal events and comparison with MC
simulations of the expected backgrounds. The fit to the
beam-constraint mass yields $4081 \pm 117$ signal events. With the total
 reconstruction efficiency of 23.3\%, the preliminary efficiency-
corrected yield of $D^0 \rightarrow \pi^0\pi^0$ based on our $\psi
(3770)$ data set is $17521 \pm 500(stat) \pm 1559(syst)$ events (
the estimation of the systematic uncertainty can be found in
Ref.~\cite{hajime}).

\begin{figure}[htbp]
  \begin{center}
    \includegraphics[width=0.3\textwidth]{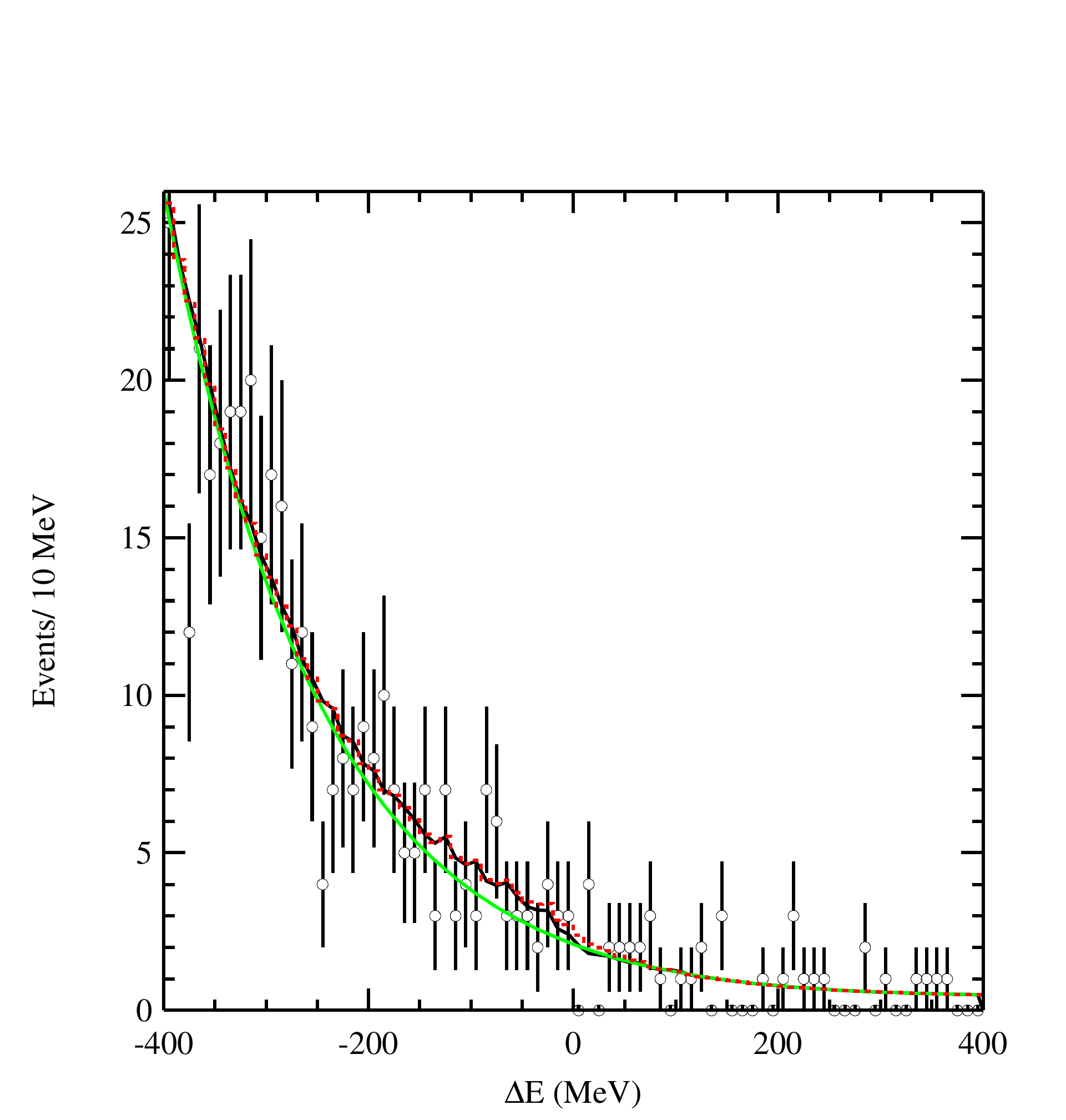}
    \caption{A fit to the $D^0 \rightarrow \gamma \gamma$
 candidate $\Delta E$ distribution based on the $\psi(3770)$ data
sample. Black points are data, solid black curve is the overall
fitted curve (signal plus backgrounds), the red-dashed curve is the
fitted total background, and the green curve is the exponential and
linear polynomials. }
    \label{fig:2g}
  \end{center}
\end{figure}

We analyze $D^0 \rightarrow \gamma \gamma$, and Fig.~\ref{fig:2g}
shows $\Delta E$ distribution based on the $\psi(3770)$ data set,
where $\Delta E$ is the energy difference  between reconstructed
energy of $D$ meson and beam energy.  The signal candidates should
be peak near zero, therefore, no significant signal events are
observed. We perform a maximum-likelihood fit to the $\Delta E$
distribution. In the fit, the signal shape is fixed by the
corresponding MC shape. The background shape consists of three
parts; MC-based shape to represent the contamination from $D^0
\rightarrow \pi^0\pi^0$ whose size is also fixed based on our own
observation; a 1st order polynominal that covers the contamination
from Bhabha events which appear smoothly over the entire $\Delta E$
spectrum; a 1st order exponential polynominal, corresponding to the
rest of the backgrounds. The fit yields $-2.9\pm7.1$ signal events.
This translates into an upper limit of 11 events at 90\% C.L. based
on the Bayesian method. Including the estimated total systematic
uncertainty, we arrive at $\mathcal{B}(D^0 \rightarrow \gamma
\gamma) /\mathcal{B}(D^0 \rightarrow \pi^0\pi^0) < 5.8 \times
10^{-3}$ at 90\% C.L.. With the known value of $\mathcal{B}(D^0
\rightarrow \pi^0\pi^0)$~\cite{pdg2012}, this corresponds to
$\mathcal{B}(D^0 \rightarrow \gamma \gamma) < 4.7 \times 10^{-6}$.

\section{Conclusion}
\label{sec:summ }

Since the start of running in 2008,  BESIII has taken about 2.9
fb$^{-1}$ of data at $\psi(3770)$.  With peak luminosity reaching
more than $6 \times 10^{32}$ cm$^{-2}s^{-1}$ (60\% of the designed
luminosity), BESIII is poised to take more data at $\psi(3770)$ and
in the higher ($D_s$) energy region. BESIII collaboration presented
the most precise measurement for $D^+ \rightarrow \mu^+\nu$ decay.
Using part of the data, BESIII has presented preliminary results of
the $D^0 \rightarrow K/\pi e \nu$ decays. Results from the full
dataset and other modes are coming in the near future.

 For the rare charm decay program, BESIII reported the search for
 $D^0 \rightarrow \gamma \gamma$ decay with single tag technique,
 and backgrounds are expected to be lower than experiment at $B$
 factories. While we are waiting for BESIII to take more data at $\sqrt{s} = 3.773$ GeV, there
is an alternate analysis approach that is unique to our data sample.
The produced $\psi(3770)$ in our sample decays into a pair of
$D^0\bar{D}^0$. Reconstructing one of the $D^0$ mesons with known
exclusive modes while searching for $D^0 \rightarrow \gamma \gamma$
in the other $D^0$ decay would yield an almost background-free
environment, except for the irreducible contamination from $D^0
\rightarrow \pi^0\pi^0$ for which we have control. Such a study is
also currently under way.  We also explore other rare charm decays
in the future with more dataset at the BESIII experiment.

\section*{Acknowledgment}
The author would like to thank his BESIII colleagues, Hajime
Muramatsu, Chunlei Liu and Gang Rong for providing many plots and
results in the paper. This work is supported in part by the National
Natural Science Foundation of China under contract No. 11125525.



%% References

\nocite{*}
\bibliographystyle{elsarticle-num}
\bibliography{martin}



\end{document}